\documentstyle[12pt]{article}
\begin{document}
\def\beq{\begin{equation}}
\def\eeq{\end{equation}}
\def\bea{\begin{eqnarray}}
\def\eea{\end{eqnarray}}
\def\ve{\vert}
\def\nnb{\nonumber}
\def\ga{\left(}
\def\dr{\right)}
\def\aga{\left\{}
\def\adr{\right\}}
\def\rar{\rightarrow}
\def\nnb{\nonumber}
\def\la{\langle}
\def\ra{\rangle}
\def\ba{\begin{array}}
\def\ea{\end{array}}

\title{ {\small { \bf THE STRONG $g_{B^{**}B\pi}$
                      COUPLING CONSTANT IN FULL $QCD$.} } }

\author{ {\small T.M.ALIEV \thanks {Permanent address: Institute of
Physics, Azerbaijanian Academy of Sciences, Baku, Azerbaijan}\,\,, and M.
SAVCI} \\ {\small Physics Department, Middle East Technical
University}\\ {\small 06531 Ankara, Turkey} }

\begin{titlepage}
\maketitle
\thispagestyle{empty}

\begin{abstract}
\baselineskip  0.7cm
To leading order in $\alpha_s$, the leading and non-leading $1/m_b$
corrections to the excited $B^{**}$ meson coupling $g_{B^{**}B\pi}$
is calculated in the framework of $QCD$ spectral
moment sum rules in the full theory. Our prediction is in good agreement
with the $light-cone~QCD$ sum rule result.

\end{abstract}
\end{titlepage}
\baselineskip  .7cm
\newpage

\setcounter{page}{1}
\section{Introduction}
At present the physics of hadrons containing heavy quarks lies in the focus
of experimentalists and theoretists. From experimental point of view it is
connected in the first hand, to the the observed exclusive $B \rar K^* \gamma$ 
\cite{R1} and
inclusive decays $B \rar X_S \gamma$ \cite{R2}, which clearly demonstrated the
existence of $FCNC$ and in turn, opens a window for the determination of the
{\it Cabibbo-Kobayashi-Maskawa} $(CKM)$ matrix elements $V_{tb}$ and
$V_{ts}$.
From theoretical point of view the reason is that the investigation of
these decays allows one to check the predictions of $CKM$ at loop level. At the
same time these decays form the basis of a  very rich {\it "laboratory
framework"} for theoretical researches. For example, in the 
infinite heavy quark mass limit $(m_Q \rar \infty)$, QCD reveals symmetries
that are not present in the finite mass theory: chiral $SU_L(3)\otimes SU_R(3)$
symmetry, spin-flavour symmetry \cite{R3}. These symmetries allow the derivation of a
simple relation among the formfactors of the different processes 
(see for review \cite{R4}) and build
an effective heavy-light Lagrangian. This effective Lagrangian widely
applied in the investigations of $B$ and $D$ meson physics. This approach
allows us to include the positive parity $(Qq)$ meson to the effective
Lagrangian according to the value of the value of the angular momentum of
the ligth degree of freedom $(s_l^P=\frac{1^+}{2},\frac{3^+}{2})$. The heavy
quark effective theory \cite{R5,R6} predicts the existence of two multiplets, the first one
containing $0^+$ and $1^+$ and the second $1^+$ , $2^+$ states mesons.

In \cite{R7} the strong coupling constant $g_{B^{**}B\pi}$,
where $B^{**}$ is the $0^+$ $p$-wave in the framework of the
{\it classical} and {\it light cone} sum rules, is investigated. 
Predictions of the both
approaches differ one another by a factor $\sim$ 1.5.
This article is devoted to the study of the large $(1/m_b)$ behavior of
the $g_{B^{**}B\pi}$ coupling constants and estimate their values in the
double momentum version of the $QCD$ sum rules. Contrary to the double
exponential version of sum rules, this version is advantageous in the
analysis of the three point function, as it prevents the blow up of the
$QCD$ series, when $m_Q$ is large. Note that the coupling constant
$g_{B^{*}B\pi}$is calculated in the framework of the above mentioned method
in \cite{R8}.

\section{Calculation of of the excited 
$g_{B^{**}B\pi}$  coupling}
We start our calculation by considering the following correlator:
\beq
V=-\int d^4x~ d^4y~e^{i(p'y-px)}\la 0 \ve
{\cal T}\aga
J_{B^{**}}(x)J_5(0)J_B(y)\adr \ve 0 \ra~,
\eeq
where the interpolating currents
\bea
J_{B^{**}}&=& m_b \bar u b~,\nnb\\
J_5&=&(m_u+m_d) \bar u\gamma_5 d~~~~~~~~~~~ and \nnb\\
J_B&=&(m_d+m_b) \bar d\gamma_5 b~,
\eea
are described $0^+$ excited $B$ meson state in the $s_l^P=\frac{1^+}{2}$
doublet, $\pi$-meson and $B$-meson states. Acoording to $QCD$ ideology this
correlator is calculated theoretically in the deep-Eucledian region and in
terms of physical states. In order perform this calculation we remind that
the correlator obeys the double dispersion relation 
\beq
V(p,p',q)=
-\frac{1}{4\pi^2}\int_{m_b^2}^{\infty}\frac{ds}{s-p^2}\int_{m_b^2}^{\infty}
\frac{ds'}{s'-p'^2}~\mbox{Im}V(s,s')~~+~~...
\eeq
Since $m_b$ is much larger than the $QCD$ scale $\Lambda$, it is natural to
use the momentum version of sum rules. Acting on both sides of Eq. (3) by
the double momentum operator,
${\ga-\frac{d}{dp^2}\dr}^n {\ga-\frac{d}{dp'^2}\dr}^n$  
over variables $p^2,p'^2$, and after differentiation setting 
$p^2=p'^2=0$, we get 
\beq
{\cal M}^{(n,n')}=-\frac{1}{4\pi^2}\int_{m_b^2}^{\infty}
\frac{ds}{s^{n+1}}
\int_{m_b^2}^{\infty}\frac{ds'}{s'^{n'+1}}~
\mbox{Im}V(s,s')~.
\eeq
The perturbative part of the spectral function can be calculated from the
Feynman diagrams by the help of the Gutkovsky rule, i.e., replacing
propagators 
\beq
\frac{1}{(p_i^2-m_i^2)} \rar -2 \pi i \delta (p_i^2-m_i^2)~.
\eeq
After carrying standard calculations for spectral density we get,
\beq
-\frac{1}{4\pi^2}\mbox{Im}V(s,s')=(m_u+m_d)m_b^3~\frac{N_c}
{8\pi^2}~ \frac{Q^2}{\lambda^{1/2}(s,s',Q^2)}~,
\eeq
where $Q^2\equiv -q^2 \geq 0$ is the pion momentum squared and
\beq
\lambda=(s+s'+Q^2)^2-4ss'~.
\eeq
Integration region defined by the following inequality:
\beq
(s-m^2_b)(s'-m^2_b)\geq Q^2m^2_b.
\eeq

The physical part of the sum rules is parametrized using the usual duality
anzats: lowest resonances + $QCD$ continuum, which usually modelled as bare
loop, starting from the thresholds $s_0$ and $s'_0$. Transferring the the
continuum contribution to the theoretical part we obtain the following $QCD$
sum rules
\beq
g_{B^{**}B\pi}~\frac{\sqrt2 m_{B^{**}}^2f_{B^{**}}}{m^{2(n+1)}_{B^{**}}}~
\frac{\sqrt2m_{B}^2f_{B}}{m^{2(n'+1)}_{B}}~
\frac{\sqrt2m^2_\pi f_\pi}{m^2_\pi+Q^2}\simeq
-\frac{1}{4\pi^2}\int_{m_b^2}^{s_0}\frac{ds}{s^{n+1}} \int_{m_b^2}^{s'_0}
\frac{ds'}{s'^{n'+1}}
{}~\mbox{Im}V(s,s'),
\eeq
In deriving the left hand side of the sum rules we use the following
definitions:
\bea
g_{B^{**}B\pi}&=&\la B(p')\pi(q)\ve B^{**}(p)\ra \nnb\\
\la 0|J_5|\pi\ra&=&\sqrt{2}f_\pi m^2_\pi  \nnb\\
\la 0|J_B|B\ra&=&\sqrt{2}f_Bm^2_B \nnb\\
\la 0|J_{B^{**}}|B^{**}\ra&=& \sqrt{2} f_{B^{**}} m_{B^{**}}^2
\eea
In the $m_b \rar \infty$ case it is convenient to work with the
non-relativistic variables $E$ and $E'$ defined as
\beq
s= (E+m_b)^2~~~~~~~\mbox{and}~~~~~~~~s'= (E'+m_b)^2,
\eeq
and introduce the new variables:
\beq
x=E-E'~~~~~~~\mbox{and}~~~~~~~~y=\frac {1}{2}(E+E')~.
\eeq
The integration region over new variables $x$ and $y$ is defined from Eq.
(8). In further analysis we shall set $n=n'=n_3$. By keeping the non-leading
terms in the expansion to the leading order in $\alpha_s$:
\bea
{\cal M}^{(n,n')}&\simeq &(m_u+m_d)~
\frac{N_c}{m_b^{4n_3}}~\frac{Q^2}{4\pi^2}\aga
\int_{-E_c^{B^{**}}}^0dx\int_{
\frac{1}{2}\sqrt{x^2+Q^2}}^{E_c^{B^{**}}+\frac{x}{2}} dy +
\int_0^{E_c^{B}}dx\int_{\frac{1}{2}\sqrt{x^2+Q^2}}^{E_c^{B}-\frac{x}{2}} dy
\adr\nnb\\
&&\Bigg{[} \frac{1}{\ga x^2+Q^2\dr^{1/2}}
\Big{[} 1-(4n_3+3) \frac{y}{m_b}
+ 2(n_3+1)(4n_3+3)\frac{y^2}{m_b^2} +\nnb\\
&&(2n_3+1) \frac{x^2}{4 m_b^2}
- \frac{Q^2}{8 m_b^2} \Big{]} \Bigg{]}~~.
\eea
Following \cite{R7,R8} , we shall use in our analysis the lowest
order expression in $\alpha_s$ for the decay constants
$f_{B^{**}}$ and $f_B$ from the 2-point momentum sum rule,

\bea
f^2_B &\simeq&
\frac{\ga E_c^{B} \dr^3}{2\pi^2}\frac{1}{m_B}
\ga \frac{m_B}{m_b}\dr^{2n_2-1} \aga
1-\frac{3}{2}(n_2+1)\frac{E^B_c}{m_b}-\frac{\pi^2}{2}~
\frac{\la \bar dd \ra }{\ga E_c^{B} \dr^3}\adr~,\nnb\\
f^2_{B^{**}}
&\simeq&\frac{\ga {E_c^{B^{**}}} \dr^3}{2\pi^2}\frac{1}{m_B^{**}}
\ga \frac{m_B^{**}}{m_b} \dr^{2n_2-1}
\aga 1-\frac{3}{2}(n_2+1)\frac{E_c^{B^{**}}}{m_b} + \frac{\pi^2}{2}~
\frac{\la \bar dd \ra }{\ga E_c^{B^{**}} \dr^3} \adr~.
\eea

From the numerical analysis of $f_B$ including the $\alpha_s$ correction one
can obtain for the effective value of $E_c^B$ \cite{R8}:
\beq
E^B_c \simeq (1.3\pm 0.1)~\mbox{GeV}~.
\eeq
Performing similiar calculation using Eq. (14) and the results cited in \cite{R9} for
the effective value $E_c^{B^{**}}$ of $f_{B^{**}}$ we get,
\beq
E_c^{B^{**}} \simeq (1.8\pm 0.1)~\mbox{GeV}~.
\eeq
Using the recently reported experimental results \cite{R10,R11} $m_{B^{**}} =
(5.732 \pm 0.002)$ MeV and $m_B=5.279$ MeV we conclude that  $m_{B^{**}}-m_B = \Delta \simeq
0.5$ GeV. From comparision of Eqs. (15) and (16) we have,
\beq
E_c^{B^{**}} \simeq E_c^B + \Delta~.
\eeq
Taking into consideration  these values of $E_c^{B^{**}}$ and $m_B^{**}$ and following \cite{R12},
we minimize the n-dependence of the results, by requiring that the leading
term is n-independent. This leads us to the following condition:
\beq
4n_3=2n_2-1~.
\eeq
From analysis of two point function it is found that $n_2\simeq4-5$
(see \cite{R13} ) .
Evaluating different integrals, the results can be written in
 the following good approximation,
\beq
g_{B^{**}B\pi}\simeq g^{LO}_{B^{**}B\pi}\aga
1+\frac{E_c^B}{m_b}\adr~,
\eeq
where:
\beq 
g^{LO}_{B^{**}B\pi}\simeq
\frac{N_c(m_u+m_d)m_B^2}{4\sqrt{2}f_\pi m^2_\pi}~~ {\cal I}_0
\eeq
and
\beq
{\cal I}_0 = \frac{Q^4}{m_B \ga E_c^B \dr^3}
\Bigg{[}\int_0^{E_c^{B^{**}}}dx\int_{\frac{1}{2}\sqrt{x^2+Q^2}}^{E_c^{B^{**}}-\frac{x}{2}} +
\int_0^{E_c^B}dx\int_{\frac{1}{2}\sqrt{x^2+Q^2}}^{E_c^B-\frac{x}{2}}\Bigg{]}
\frac{dy}{\ga x^2+Q^2\dr^{1/2}}~.
\eeq
In the region $1\leq Q^2  \leq 4$ this integral has a maximum at
$Q^2\simeq 2.5~ \mbox{GeV}^2$. Using this value of $Q^2$,
numerical integration of $~{\cal I}_0~$ for different values of $E_c^B$ can be
parametrized by following interpolating formula
\beq
{\cal I}_0=( 0.175 \pm 0.001)E_c~.
\eeq
Using
\beq
E_c^B=E_c^{\infty}\ga 1-\frac{E_c^B}{2m_b}\dr ,~~~\mbox{(see for example}~~
\cite{R8}~ )~,
\eeq
we finally get
\beq
g_{B^{**}B\pi} \simeq \frac{3 m_B^2}{4\sqrt{2}f_\pi} g^{\infty} \aga
1+\frac{E_c^B}{2 m_b}\adr~.
\eeq
Using the result $g^{\infty}=(0.15 \pm 0.03)$, that is given in \cite{R8} 
and taking into consideration the $\frac {1}{m_b}$ correction
for the $physical
~B-meson$ (which is of the order of $\pm28\%$ at the $b$-mass),
we get,
\beq
g_{B^{**}B\pi} = 24 \pm 7~~GeV~,
\eeq
where the error in the above result takes into account the effect of
radiative corrections to $f_B$.
As mentioned in Introduction, the value of the strong coupling constant
$g_{B^{**}B\pi}$ that is calculated in \cite{R7} within the framework of the
{\it classical} and {\it light-cone sum rule} techniques are , 
\bea
g_{B^{**}B\pi}&=&13.3 \pm 4.8~~GeV,\nnb \\
g_{B^{**}B\pi}&=&21  \pm 7~~GeV ,
\eea
respectively. From comparision of Eqs. (25) and (26) we see that 
our prediction on $g_{B^{**}B\pi}$ is in
good agrrement with the {\it light-cone} $QCD$ sum rule prediction on the
same quantity.

\end{document}